\documentclass[prb,onecolumn,showpacs,superscriptaddress,preprintnumbers,amssymb]{revtex4}
\usepackage{graphicx}% Include figure files
\usepackage{dcolumn}% Align table columns on decimal point
\usepackage{bm}% bold math
\usepackage{wasysym}
\newcommand{\beq}{\begin{equation}}
\newcommand{\eeq}{\end{equation}}
\newcommand{\beqn}{\begin{eqnarray}}
\newcommand{\eeqn}{\end{eqnarray}}
\newcommand{\nn}{\nonumber\\}

\newcommand{\sgn}{{\rm sgn}}

\def\F{\mathcal{F}}
\def\sgn{{\rm sgn}}
\def\V{\mathcal{V}}
\def\U{\mathcal{U}}

\begin{document}

\title{A model for continuous thermal Metal to Insulator Transition}

\author{Chao-Ming Jian}
\affiliation{Kavli Institute of Theoretical Physics, Santa
Barbara, CA 93106, USA}
\affiliation{
Station Q, Microsoft Research, Santa Barbara, California 93106-6105, USA}
\author{Zhen Bi}
\affiliation{Department of Physics, University of California,
Santa Barbara, CA 93106, USA}
\author{Cenke Xu}
\affiliation{Department of Physics, University of California,
Santa Barbara, CA 93106, USA}

\date{\today}
\begin{abstract}

We propose a $d-$dimensional interacting Majorana fermion model
with quenched disorder, which gives us a continuous quantum phase
transition between a diffusive thermal metal phase with a finite
entropy density to an insulator phase with zero entropy density.
This model is based on coupled Sachdev-Ye-Kitaev model clusters,
and hence has a controlled large-$N$ limit. The metal-insulator
transition is accompanied by a spontaneous time-reversal symmetry
breaking. We perform controlled calculations to show that the
energy diffusion constant jumps to zero discontinuously at the
metal-insulator transition, while the time-reversal symmetry
breaking order parameter increases continuously.

\end{abstract}

\maketitle

\section{Introduction}

The metal to insulator transition (MIT) is a fairly old subject.
In a noninteracting fermion system with quenched disorder, MIT can
occur by tuning the Fermi surface across the mobility edge of the
single particle spectrum~\cite{anderson}; in a system with
interacting electrons, MIT can be driven by the competition
between interaction and kinetic energy, $i.e.$ the so-called Mott
transition~\cite{mott}, and the Mott insulator phase often
develops a Landau order parameter which spontaneously breaks
certain symmetry of the system, though a more interesting
possibility is that the Mott insulator phase close to the Mott
transition is a spin liquid phase without any Landau order
parameter~\cite{sl1,sl2,spinqliquid1,spinliquid2,spinliquid3,mottsenthil,xudid}.

In recent years, MIT between highly excited states with finite
energy density have attracted enormous interests and efforts. This
type of MIT is driven by the interplay between interaction and
quenched disorder. On one side of the transition, the state obeys
the eigenstate thermalization hypothesis (ETH)~\cite{eth1,eth2},
and the many-body eigenstate is extended (either in the real space
or the fock space) with a volume-law entanglement entropy, while
the other side of the transition is a many-body localized (MBL)
state~\cite{mblanderson,mblpolyakov,mblaltshuler,mblhuse1,mblhuse2}
which is a product of eigenstates of localized conserved
quantities~\cite{mblabanin,mblhuse3} and hence only possesses
boundary-law entanglement entropy. Great numerical and analytical
efforts have been made towards understanding the nature of the
ETH-MBL
transition~\cite{mblaltshuler,mblhuse2,transitiongarel,transitionpollman,
transitionhuse,transitionaltman,transitionaltman2,transitionpotter,
transitionabanin2,transitionalet,transitionsantos,transitionsingh,muller,
moore,huse5,huse6,transitionpotter2,grover2014}. Due to the
difficulty of directly studying the ETH-MBL transition in a
generic nonintegrable many-body Hamiltonian, studies based on
many-body wave functions instead of Hamiltonians have also been
pursued~\cite{wavefunction1,grover2014b}. However, an explicit
Hamiltonian with an ETH-MBL transition that can be studied using
controlled analytical method is still demanded.

In this work we construct a many-body Hamiltonian which gives us a
quantum MIT at its ground state. The basic degrees of freedom in
our model are Majorana fermions, which only transport energy
rather than electric charge. The thermal metallic phase of our
model has a finite entropy density and energy diffusion constant;
while the insulator phase has zero entropy density and diffusion.
%The ground state MIT of our model to some extent
%resembles the ETH-MBL transition at finite energy density, as a
%MBL state at finite energy density can be viewed as the ground
%state of a local Hamiltonian, since it only has boundary-law
%entanglement entropy. Also, the MIT in our model is accompanied by
%a spontaneous time-reversal symmetry breaking, which is analogous
%to the conventional Mott transition.
The main advantage of our model is that it is accessible
analytically with controlled methods, if the fermion flavor number
$N$ on every site (cluster) is taken to be large. If we take the
large$-N$ limit, and the infrared limit, the diffusion constant
would jump discontinuously to zero at the transition.

This paper is organized as follows: in section II we will first
derive the saddle point equations of our model, and the mean field
solution of the saddle point equations demonstrates that a quantum
phase transition occurs by tuning one parameter in our model; in
section III we perform a large$-q$ calculation to demonstrate that
the transition we construct is between a thermal metal and
insulator with finite and zero energy diffusion constants
respectively; in section IV we use the effective action of quasi
Goldstone modes and scaling argument to verify the results in
section III.

\section{The Model and Saddle Point Solutions}

\begin{figure}[tbp]
\begin{center}
\includegraphics[width=320pt]{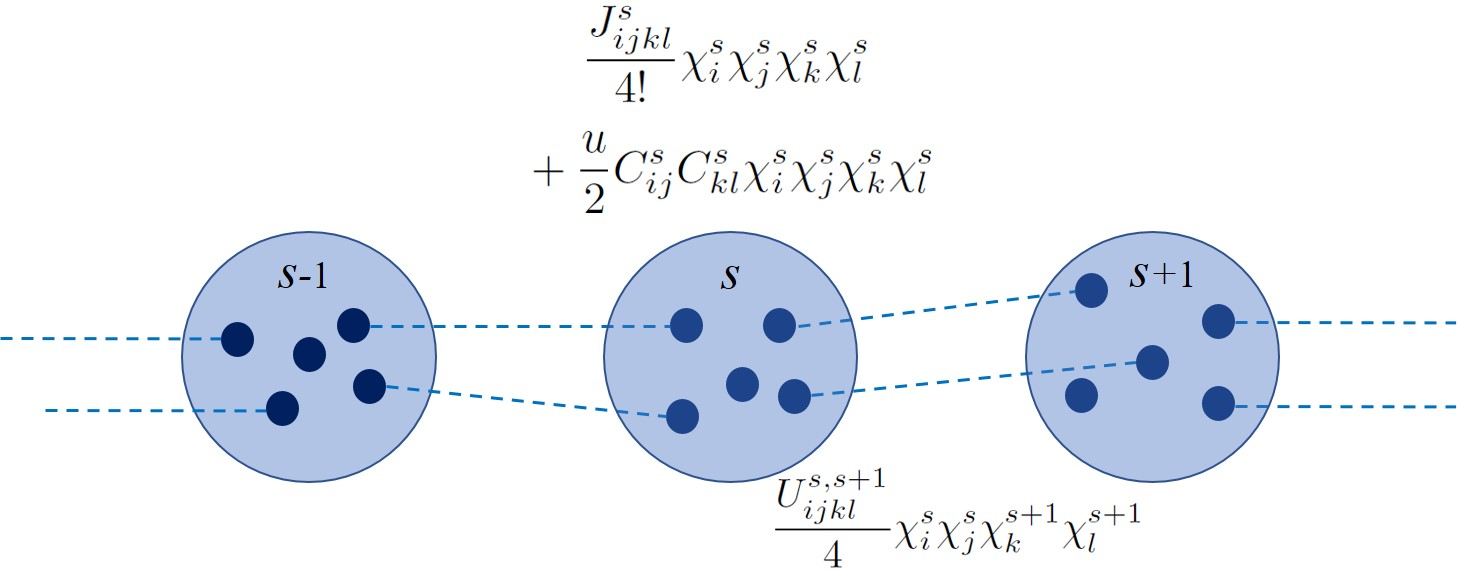}
\caption{The schematic structure of our model Eq.~\ref{chain}.}
\label{chainfig}
\end{center}
\end{figure}

We first consider the following Hamiltonian for a one dimensional
chain (it is trivial to generalize our model to higher dimensions,
and the calculations and conclusions in this work are insensitive
to the spatial dimension), with a large number ($N$) of Majorana
fermions defined on each site (cluster) $s$: \beqn \label{chain}
H_{\mathrm{chain}} &=& \sum_{s=1}^L \frac{J^s_{ijkl}}{4!}
\chi^s_i\chi^s_j\chi^s_k\chi^s_l + \frac{u}{2} C^{s}_{ij}
C^{s}_{kl} \chi^s_i\chi^{s}_j\chi^{s}_k\chi^{s}_l +
\frac{U^{s,s+1}_{ijkl}}{4} \chi^s_i \chi^s_j \chi^{s+1}_k
\chi^{s+1}_{l}. \eeqn $J^s$, $C^s$ and $U^{s,s+1}$ are all
independent random variables which have zero mean and obey the
Gaussian distribution: \beqn \label{distribution} N^3
\overline{(J^{s}_{ijkl})^2} = 3!J^2, \ \ \ \
N^2\overline{(C^{s}_{ij})^2} = C^2, \ \ \ \ N^3
\overline{(U^{s,s+1}_{ijkl})^2} = U^2. \eeqn The structure of this
model is schematically shown in Fig.~\ref{chainfig}

Several limits of this model have been understood:

{\it 1.} When $U = 0$, this model becomes decoupled sites with the
first two terms of the Hamiltonian. The first term is the ordinary
Sachdev-Ye-Kitaev model (SYK$_4$) with Majorana
fermions~\cite{SachdevYe1993,Kitaev2015,MaldacenaStanford2016},
whose ground state in the large$-N$ limit is a non-fermi liquid
with power-law fermion Green's function after disorder average,
and the scaling dimension of fermion $\chi_j$ is $[\chi_j] =
1/4$~\footnote{The power-law fermion Green's function has the
SL(2,R) ``conformal" symmetry in the infrared, which spontaneously
breaks the larger reparametrization symmetry of the
model~\cite{Kitaev2015,MaldacenaStanford2016} (if we ignore the
$\partial_\tau$ term of the Lagrangian).}. It was shown
recently~\cite{SYKinstable} that if we treat the $u$ term as a
perturbation on the non-fermi liquid state described by the
ordinary SYK$_4$ model (the first term of Eq.~\ref{chain}), it is
marginally relevant (irrelevant) when $u > 0$ ($u < 0$). And for
$u > 0$, the $u$ term will lead to spontaneous time-reversal
symmetry breaking and generate a SYK$_2$ term with a random
two-body interaction, at an exponentially low energy scale. This
seemingly zero dimensional system is capable of having its own
phase transition because it has infinite degrees of freedom in the
large$-N$ limit. The spontaneously generated SYK$_2$ term will
dominate the physics in the infrared, and the SYK$_4$ term becomes
irrelevant.

{\it 2.} When $u = 0$, this model becomes the coupled SYK clusters
studied in Ref.~\onlinecite{Gu2016}. A complex fermion version of
this case was studied later~\cite{Davison2016}. Both the on-site
$J^s$ term and the inter-cluster $U^{s,s+1}$ term contribute on
the equal footing. The entire $1d$ system is a diffusive thermal
metal, with a finite energy diffusion constant, and a ``butterfly
velocity" due to the chaotic nature of the SYK$_4$ model.

In this work we treat the onsite $u$ term as a perturbation on the
coupled SYK cluster chain. It is straightforward to show that, the
renormalization group flow of the $u$ term is almost identical to
the single cluster case discussed in
Ref.~\onlinecite{SYKinstable}, $i.e.$ when $u < 0$, it can be
ignored in the infrared limit because it is marginally irrelevant;
while when $u > 0$, it becomes marginally relevant. The saddle
point equation that we will discuss next demonstrates that in the
latter case the system again spontaneously generates an onsite
SYK$_2$ random two-body interaction term, which dominates all the
low energy physics. If we start with an onsite SYK$_2$ term, then
both the $J^s$ and $U^{s,s+1}$ terms are irrelevant, and the
system becomes decoupled clusters with zero entropy density. Since
the inter-cluster coupling will renormalize to zero in the
infrared limit, this system is expected to be an insulator for
$u>0$. Thus tuning $u$ from negative to positive value will lead
to a quantum phase transition between a thermal metal with finite
entropy density and energy diffusion, to an insulator with zero
entropy density and diffusion constant.

To demonstrate the effect mentioned above, we follow
Ref.~\onlinecite{SYKinstable} and perform a Hubbard-Stratonovich
transformation for the $u$-term. The action for the chain becomes:
\beq \mathcal{S} = \sum_{s=1}^L\int d\tau
\frac{1}{2}\chi_i^s\partial_\tau\chi_i^s+\frac{J^s_{ijkl}}{4!}\chi^s_i\chi^s_j\chi^s_k\chi^s_l+\frac{U^{s,s+1}_{ijkl}}{4}
\chi^s_i \chi^s_j
\chi^{s+1}_k\chi^{s+1}_{l}+\frac{u}{2}b^{s}b^{s}-iuC^{s}_{jk}b^{s}\chi_j^s\chi_k^{s}.
\label{Eq:chain_action}
\eeq

With the normalizations in Eq.(\ref{distribution}), we can write
down the disorder averaged theory. Under the replica diagonal
assumption, the action reads \beqn \mathcal{S} &=&\sum_{s=1}^L\int
d\tau \left(\frac{1}{2}\chi_i^s\partial_\tau\chi_i^s +
\frac{u}{2}(b^{s})^2\right)-\sum_{s=1}^Lu^2\frac{C^2}{N^2}\int
d\tau_1 d\tau_2 \
b^{s}(\tau_1)b^{s}(\tau_2)(\chi_j^s(\tau_1)\chi_j^s(\tau_2))^2 \nn
&&-\sum_{s=1}^L\frac{1}{8N^3}\int
d\tau_1d\tau_2\left[J^2(\chi_i^s(\tau_1)\chi_i^s(\tau_2))^4
+U^2(\chi_i^s(\tau_1)\chi_i^s(\tau_2))^2(\chi_j^{s+1}(\tau_1)\chi_j^{s+1}(\tau_2))^2\right]
\eeqn Following Ref.~\onlinecite{MaldacenaStanford2016}, we
proceed by introducing the fermion Green's function and
self-energy on every site $s$: \beqn \mathcal{S}_{eff} &=&
\sum_{s=1}^L\int d\tau_1d\tau_2
\left(\frac{1}{2}\chi_i^s(\tau_1)(\delta(\tau_1-\tau_2)\partial_{\tau_2}-\Sigma^s(\tau_1,\tau_2))
\chi_i^s(\tau_2)
+\frac{u}{2}b^{s}(\tau_1)\delta(\tau_1-\tau_2)b^s(\tau_2)\right)
\nn && -\sum_{s=1}^Lu^2C^2\int d\tau_1 d\tau_2 \
b^{s}(\tau_1)b^{s}(\tau_2)(G^s(\tau_1,\tau_2))^2+\frac{N}{2}\sum_{s=1}^L\int
d\tau_1d\tau_2 \ \Sigma^s(\tau_1,\tau_2)G^s(\tau_1,\tau_2)\nn
&&-\frac{N}{2}\sum_{s=1}^L\int
d\tau_1d\tau_2\left[\frac{J^2}{4}(G^s(\tau_1,\tau_2))^4+\frac{U^2}{4}(G^s(\tau_1,\tau_2))^2(G^{s+1}(\tau_1,\tau_2))^2\right]
\eeqn According to previous studies of instabilities of SYK
model~\cite{SYKinstable}, for $u > 0$, the bosonic field $b$ on
each site will develop long-range correlation, thus as a mean
field approximation we can replace the composite field
$b^s(\tau_1)b^s(\tau_2)$ by its expectation value $\langle
b^s(\tau_1)b^s(\tau_2)\rangle= N (w^s)^2$, where $w^s$ is an
$O(1)$ number and should be self-consistently determined by the
saddle point equations. The large-$N$ saddle-point-mean-field
equations for the above action are: \beqn
&&\tilde{G}^s(i\omega_n)^{-1}=-i\omega_n-\tilde{\Sigma}^s(i\omega_n)
\nn \nn
&&\Sigma^s(\tau)=4u^2C^2(w^s)^2G^s(\tau)+G^s(\tau)\left[J^2G^s(\tau)^2+\frac{U^2}{2}(G^{s-1}(\tau)^2+G^{s+1}(\tau)^2)\right]
\nn \nn &&\int_0^\beta d\tau
\left(uC^2G^s(\tau)^2-\frac{1}{2}\delta(\tau)\right)uw^s=0 \eeqn
If we only consider translational invariant solutions, we can drop
the site indices, and the saddle-point-mean-field equations
become: \beqn
&&\tilde{G}(i\omega_n)^{-1}=-i\omega_n-\tilde{\Sigma}(i\omega_n)\nn
\nn &&\Sigma(\tau)=4u^2C^2w^2G(\tau)+(J^2+U^2)G(\tau)^3 \nn \nn
&&\int_0^\beta d\tau
\left(uC^2G(\tau)^2-\frac{1}{2}\delta(\tau)\right)uw=0 \eeqn

We arrive practically the same saddle-point-mean-field equations
as the single site model studied previously in
Ref.~\onlinecite{SYKinstable}. For $u < 0$, the only solution for
the saddle point equations has $w = 0$. In this limit, the model
resembles the feature of the SYK chain model previously studied by
Gu $et. al.$\cite{Gu2016}, and the system is a diffusive metal.
For $u > 0$, the solution of the saddle point equations gives $w$
a non-zero value, which for small positive $u$ is approximately
given by \beqn w \sim \frac{(J^2 + U^2)^{1/2}}{u C} \exp\left( -
\frac{\sqrt{\pi}( J^2 + U^2)^{1/2}}{4u C^2} \right); \eeqn and the
system spontaneously breaks the time reversal symmetry.

In the time-reversal symmetry breaking phase, the first term
$4u^2C^2w^2G(\tau)$ in the self energy dominates at low energy. In
the deep infrared, the fermions on each site behave identically as
the SYK$_2$ model and their scaling dimension $[\chi]=1/2$.
Therefore, the $J^s$ and $U^{s,s+1}$ interactions are irrelevant
in the infrared and the chain becomes an array of isolated
decoupled SYK$_2$ clusters. We expect the chain becomes an
insulator in this limit, which will be verified with a large-$q$
expansion method in the following section. Thus tuning $u$ from
negative to positive drives a continuous phase transition between
a diffusive thermal metal and an insulator phase. Notice that $w$
increases continuously from zero starting $u = 0$, but it is
exponentially small at small $u$, which is analogous to the BCS
instability of the fermi liquid, and the Kondo effect, and it is a
sign of the marginally-relevant nature of $u$~\cite{SYKinstable}.

\section{Large $q$ analysis}

\label{Sec:large_q} For $u > 0$, when $w$ develops a non-trivial
vacuum expectation value, the system is effectively described by
the Hamiltonian $H_{\text{chain}}$ but with the second term
replaced by on-site SYK$_2$ terms. This replacement is equivalent
to, in the path integral formalism, setting $b^s$ to their vacuum
expectation values in Eq. \ref{Eq:chain_action} and neglecting
their fluctuations which should be unimportant at low energies.
Further, to justify that system becomes an insulator in the IR at
zero temperature, it is enough to just consider the effective
on-site SYK$_2$ couplings and the $U$ interactions (because the
$J$ does not couple different sites). Therefore, we consider the
effective model: \beqn H_{\text{eff}} = \sum_s \sum_{1\leq i_1 <
i_2 \leq N} iV^s_{i_1 i_2 } \chi^s_{i_1} \chi^s_{i_2} + \sum_s
\sum_{ 1\leq i_1 < i_2  \leq N  \atop 1\leq i_{3} < i_{4} \leq N}
U^{s,s+1}_{i_1 i_2 i_3 i_4} \chi^s_{i_1} \chi^s_{i_2}
\chi^{s+1}_{i_{3}} \chi^{s+1}_{i_{4}}, \label{Eq:Ham_Cluster_2_4}
\eeqn where $V^s_{i_1 i_2 }$ are independent Gaussian random
variables which correspond to $u \langle b_s \rangle C_{i_1 i_2}$
in the original model. We also redefined the coefficient of the
$U$ term for later convenience. In fact, $H_{\text{eff}}$ belongs
to a large family of models: \beqn H_{q}= (i)^{q/2} \sum_s
\sum_{1\leq i_1 < i_2 ... < i_q \leq N} V^s_{i_1 i_2 ...i_q}
\chi^s_{i_1} \chi^s_{i_2} ... \chi^s_{i_q} + (i)^{q} \sum_s \sum_{
1\leq i_1 < i_2 ... < i_{q} \leq N \atop 1\leq i_{q+1} < i_{q+2}
... < i_{2q} \leq N} U^{s,s+1}_{i_1 i_2 ...i_{2q}} \chi^s_{i_1}
\chi^s_{i_2} ... \chi^s_{i_q} \chi^{s+1}_{i_{q+1}}
\chi^{s+1}_{i_{q+2}} ... \chi^{s+1}_{i_{2q}},
\label{Eq:Ham_Clust_q_2q} \eeqn where we consider SYK$_q$ on-site
coupling and SYK$_{2q}$ inter-cluster coupling with an even
integer $q$. The couplings $V^s_{i_1 i_2 ...i_q}$ and
$U^{s,s+1}_{i_1 i_2 ...i_{2q}}$ are independent random variables
with zero mean and Gaussian distribution: \beqn
\overline{\left(V^s_{i_1 i_2 ...i_q}\right)^2} = \frac{V^2 (q-1)!
}{N^{q-1}} \ , \ \ \ \ \ \ \overline{\left(U^{s,s+1}_{i_1 i_2
...i_{2q}}\right)^2} = \frac{U^2 (q-1)! \ q!}{N^{2q-1}}. \eeqn The
effective Hamiltonian $H_{\text{eff}}$ can be recovered by
choosing $q=2$. In fact, we expect $H_q$ for all $q$ to share the
same feature that the model effectively becomes isolated SYK$_q$
clusters in the IR, where the inter cluster couplings (the $U$
terms) are irrelevant. Hence, the system should be an insulator at
low energy. To justify this expectation, we will solve this family
of models using the large-$q$ expansion developed in Ref.
\onlinecite{MaldacenaStanford2016}. It was shown in previous
studies\cite{MaldacenaStanford2016, Fu2016}, this large-$q$
expansion method is capable of capturing the thermodynamical
properties of the SYK model and its generalizations at low
energies.

We first investigate the 2-point Green's function of this model.
The saddle point equations of the Green's function $G^s(\tau) =
\frac{1}{N} \sum_i \langle \chi^s_i (\tau) \chi^s_i(0) \rangle$ of this model are
given by \beqn && G^s(i\omega)^{-1} = -i\omega - \Sigma^s(i\omega)
\label{Eq:Saddle_Pt_GS} \\
 && \Sigma^s(\tau) =
G^s(\tau)^{q-1} \Big[ V^2  + U^2 G^{s-1}(\tau)^{q} + U^2
G^{s+1}(\tau)^{q} \Big] \label{Eq:Saddle_Pt_SG} \eeqn In the large
$q$ limit, we can expand the $G^s(\tau)$ as a function of $1/q$:
\beqn & G^s (\tau) = \frac{1}{2} \sgn(\tau) \Big[ 1 + \frac{1}{q}
g^s (\tau) + ...\Big], \eeqn where ``$...$" represents higher
order terms in $1/q$ which will be neglected in the following.
Notice that when $q\rightarrow \infty$, the Green's function
becomes that of free fermions without any dispersion. In the
following, we will obtain an analytical solution of $g^s(\tau)$,
that is the $1/q$ correction to the Green's function. At finite
temperature $\beta$, $g^s(\tau)$ has to satisfy the boundary
conditions $ g^s(\tau = 0) = g^s(\tau = \beta) = 0$. Given this
expansion of $G^s (\tau) $, we can express the self energy via Eq.
\ref{Eq:Saddle_Pt_SG} as \beqn \Sigma^s(\tau) =
\frac{\sgn(\tau)}{q} e^{g^s(\tau)} \Big[ \mathcal{V}^2  +
\frac{\mathcal{U}^2}{2} e^{g^{s-1}(\tau)} +
\frac{\mathcal{U}^2}{2}  e^{g^{s+1}(\tau)}\Big].
\label{Eq:Saddle_Point_Ansatz} \eeqn where \beqn \mathcal{V}^2
\equiv  \frac{q }{2^{q-1}} V^2,~~~~ \mathcal{U}^2 \equiv
\frac{2q}{2^{2q-1}}  U^2 \eeqn are the coupling constants kept
fixed in the large $q$ limit. Plugging this expression back to the
saddle point equation Eq. \ref{Eq:Saddle_Pt_GS}, we can obtain the
translationally invariant solution: \beqn
 e^{g_s(\tau)} = \frac{1}{\frac{2 \beta^2 \mathcal{V}^2}{\pi^2 v^2} + \sqrt{\frac{4 \beta^4 \mathcal{ V}^4}{ \pi^4 v^4} + \frac{2 \beta^2 \mathcal{U}^2}{\pi^2 v^2}} \cos \left( \pi v \Big(\frac{|\tau|}{\beta}-\frac{1}{2}\Big)\right)}.
\label{Eq:Self_Energy} \eeqn Here the parameter $v \in (1, 2)$
given by the relation \beqn \cos \left(\frac{\pi v}{2}\right) = -
\frac{1- \frac{v^2 \pi^2}{2\beta^2 \mathcal{V}^2}}{\sqrt{1+
\frac{\mathcal{U}^2}{\mathcal{V}^2} \frac{v^2 \pi^2}{2\beta^2
\mathcal{V}^2}}} \eeqn ensures the boundary conditions of
$g^s(\tau)$. By taking the low temperature limit $\beta
\rightarrow \infty$, we can expand $v$ in terms of $(\beta
\mathcal{V})^{-1}$: \beqn & v = 2 -  \sqrt{ 1+
\frac{\mathcal{U}^2}{ 2 \mathcal{V}^2}  } \frac{4 }{\beta
\mathcal{V}} + O\left(\frac{1 }{\beta^2 \mathcal{V}^2}\right).
%\\
% \cos \left(\frac{\pi v}{2}\right) = -1 +\left( 1+ \frac{\mathcal{U}^2}{2 \mathcal{V}^2}  \right) \frac{v^2 \pi^2}{2\beta^2 \mathcal{V}^2} \\
% \sin \left(\frac{\pi v}{2}\right) = \sqrt{ 1+ \frac{\mathcal{U}^2}{ 2 \mathcal{V}^2}  } \frac{v \pi}{\beta \mathcal{V}},
\eeqn More importantly, $e^{g_s(\tau)} $ also has a well-defined
low temperature limit: \beqn e^{g_s(\tau)} = \frac{1}{1+ 2
\sqrt{1+ \frac{\mathcal{U}^2 }{2 \mathcal{V}^2}} \mathcal{V}|\tau|
+ \mathcal{V}^2 \tau^2 } . \eeqn As we can see from this
expression, in the long time limit, $e^{g_s(\tau)}$ behaves as
$(\mathcal{V}\tau)^{-2}$, which is independent from the coupling
constant $\mathcal{U}$. This agrees with the expectation that the
low energy physics is governed by the on-site SYK$_q$ couplings
and the inter cluster couplings become irrelevant at IR.

To verify the insulating behavior of $H_q$ in the IR, we need to
show that all the lower energy modes are localized in space. As is
pointed out by Ref. \onlinecite{MaldacenaStanford2016,Gu2016}, the
propagation of low energy modes  are all captured by the 4-point
function \beqn \frac{1}{N} \F^{ss'}(\tau_1, \tau_2, \tau_3,
\tau_4) \equiv \frac{1}{N^2} \sum^N_{i,j=1} \langle T
\big(\chi^s_i(\tau_1) \chi^s_i(\tau_2) \chi^{s'}_j(\tau_3)
\chi^{s'}_j(\tau_4)\big) \rangle - \delta_{ss'} G^s(\tau_{12})
G^s(\tau_{34}), \eeqn where we've used the notation $\tau_{ij}
\equiv \tau_i -\tau_j$ and $G^s(\tau)$ takes the form of the
solution of the saddle point equation. Since the saddle point
solution $G^s$ is translational invariant, site index $s$ on $G^s$
will be ignored hereafter. Physically, $\F^{ss'}(\tau_1, \tau_2,
\tau_3, \tau_4) $ captures the propagation, from site $s$ to site
$s'$, of all the modes created by a pair of fermion operators (at
the same site but at different times). In the large $N$ limit,
$\F^{ss'}(\tau_1, \tau_2, \tau_3, \tau_4)$ can be calculated via
the ladder diagrams\cite{MaldacenaStanford2016,Gu2016}. The
contribution of the $n$-rung ladder is denoted as $\F^{(n)}$.
First of all, we have \beqn \F^{ss'(0)}(\tau_1, \tau_2, \tau_3,
\tau_4) = \delta_{ss'}\Big( - G(\tau_{13}) G(\tau_{24}) +
G(\tau_{14}) G(\tau_{23}) \Big). \eeqn The contribution of the
$n$-rung ladder can be obtain from that of the $(n-1)$-rung
ladder: \beqn \F^{ss'(n)} (\tau_1, \tau_2, \tau_3, \tau_4) =
\sum_{s''} \int^{\beta}_0 d\tau d\tau' K^{ss''} (\tau_1, \tau_2,
\tau, \tau') \F^{s''s'(n-1)}(\tau, \tau', \tau_3, \tau_4).
\label{Eq:Ladder_Iteration} \eeqn with the kernel $K$ given by
\beqn && K_{ss'} (\tau_1, \tau_2, \tau_3, \tau_4)
\nonumber \\
&& ~~ = - G(\tau_{13}) G(\tau_{24})  \Big[  \Big( (q-1)  V^2
G(\tau_{34})^{q-2} +  2(q-1) U^2 G(\tau_{34})^{2q-2} \Big)
\delta_{ss'} + q U^2 G(\tau_{34})^{2q-2}  (\delta_{s+1,s'} +
\delta_{s-1,s'})  \Big]. \eeqn It is useful to think about
$K^{ss'} (\tau_1, \tau_2, \tau_3, \tau_4) $ as a matrix with the
left index given by $s$, $\tau_1$ and $\tau_2$, and the right
index given by $s'$, $\tau_3$ and $\tau_4$. From this perspective,
Eq. \ref{Eq:Ladder_Iteration} can be viewed as a matrix
multiplication. Then, the 4-point function, as a summation of all
the ladder diagrams, can be written as \beqn \F = \frac{1}{1-K}
\F^{(0)}. \eeqn The form of the ladder diagram suggested that
$\frac{1}{1-K}$ should be viewed as the propagator whose
eigenvectors correspond to different orthogonal modes in the
system. These orthogonal modes are also eigenvectors of $K$, among
which the ones with eigenvalues close to $1$ correspond to the low
energy modes of the system. The kernel $K$ is diagonal in momentum
space: \beqn K_{p} (\tau_1, \tau_2, \tau_3, \tau_4)  =  -
\frac{\sgn(\tau_{13})}{2} \frac{\sgn(\tau_{24})}{2} \Big[   2\V^2
e^{g(\tau_{34})}+ 2\U^2 e^{2g(\tau_{34})} (1+\cos(p)) \Big], \eeqn
where $ K_{p} \equiv \sum_s K^{ss'} e^{ip(s-s')}$ is the
real-space Fourier transform of $K_{ss'}$ and $p$ denotes the
momentum in real space. Here, we've plugged in the saddle point
solution of $G(\tau)$ and omitted the higher order terms in $1/q$.
The eigenvectors $\Psi_p(\tau_1, \tau_2)$ and eigenvalues $k_p$ of
$K_p$ are given by the equation (to the leading order in $1/q$):
\beqn k_p \Psi_p(\tau_1, \tau_2) = - \int d \tau_3 d\tau_4
\frac{\sgn(\tau_{13})}{2} \frac{\sgn(\tau_{24})}{2} \Big[   2\V^2
e^{g(\tau_{34})}+ 2\U^2 e^{2g(\tau_{34})} (1+\cos(p))
\Big]\Psi_p(\tau_3, \tau_4). \eeqn Following Ref.
\onlinecite{MaldacenaStanford2016}, we can take the derivative
$\partial_{\tau_1}\partial_{\tau_2}$ on both sides of the equation
and obtain \beqn k_p  \partial_{\tau_1}\partial_{\tau_2}
\Psi_p(\tau_1, \tau_2) = -  \Big[   2\V^2 e^{g(\tau_{12})}+ 2\U^2
e^{2g(\tau_{12})} (1+\cos(p)) \Big] \Psi_p(\tau_1, \tau_2).
\label{Eq:Eigen_Diff} \eeqn Since this equation is invariant under
simultaneous translation of the two time variables $\tau_1$ and
$\tau_2$, we can directly work with the ansatz $\Psi_{p,n}(\tau_1,
\tau_2) = e^{-i \frac{2\pi n}{\beta} \frac{\tau_1 + \tau_2}{2}
}\psi_{p,n}(\tau_{12})$ with integer $n$. Further, by introducing
the variables \beqn & \rho = \cos \left( \pi v
\Big(\frac{|\tau_{12}|}{\beta}-\frac{1}{2}\Big)\right)
~~\text{and}~~ \alpha = \sqrt{1+ \frac{v^2}{8}  \frac{\U^2}{\V^2}
\frac{4\pi^2}{\beta^2\V^2}  }, \eeqn we can simplify Eq.
\ref{Eq:Eigen_Diff} to \beqn \Bigg[ v^2\left((1-\rho^2)
\frac{d^2}{d \rho^2} -\rho \frac{d}{d \rho} \right) +\frac{n^2}{4}
- \frac{v^2}{4 k_{p,n}} \left( \frac{1}{1+\alpha \rho}
+\frac{(\alpha^2 -1)(1+\cos p)}{(1+\alpha \rho)^2}  \right) \Bigg]
\psi_{p,n} = 0. \eeqn This equation can be studied perturbatively
by treating $\alpha$ and $v$ as independent variables and by
viewing $\alpha-1$ as a small parameter. At the zeroth order,
namely $\alpha = 1$, the dependence of this equation on the
momentum $p$ simply drops out. We recover the same differential
equation studied in Ref. \onlinecite{MaldacenaStanford2016} which
gives rise to the following eigenvalues \beqn k_{n,p}^{(0)} & = 1
-\frac{3|n|}{2}\Big(1-\frac{v}{2}\Big) +
O\Big(\Big(1-\frac{v}{2}\Big)^2\Big) \nonumber \nn \nn & = 1 -
\frac{3|n|}{\beta \V} \sqrt{ 1+ \frac{\mathcal{U}^2}{ 2
\mathcal{V}^2}  } + O\Big(\frac{1}{\beta^2 \V^2}\Big). \eeqn where
the superscript ``$^{(0)}$" represents the zeroth order in
$\alpha-1$. Here, we've only listed the eigenvalues that are close
to 1 since we are only interested in the low energy modes in the
theory. The momentum dependence of the eigenvalue $k_{n,p}$ comes
into the first order correction in $(\alpha-1)$. However, since
$(\alpha-1) \sim \frac{1}{\beta^2 \V^2}$ in the low temperature
limit $\beta\V \rightarrow \infty$, even including the first order
correction, we still have \beqn k_{n,p}^{(1)} = 1 -
\frac{3|n|}{\beta \V} \sqrt{ 1+ \frac{\mathcal{U}^2}{ 2
\mathcal{V}^2}  } + O\Big(\frac{1}{\beta^2 \V^2}\Big),
\label{Eq:k1} \eeqn where the $O\Big(\frac{1}{\beta^2 \V^2}\Big)$
part also depends on the momentum $p$. When we view $(1-
k_{n,p})^{-1}$ as the propagator of low energy modes, we should
rewrite $n$ using the Matsubara frequency $\omega_n = \frac{2\pi
n}{\beta}$. It is immediately clear that the leading term in the
propagator $(1- k_{n,p})^{-1}$ behaves as $\left(
\frac{3\omega_n}{2\pi\V} \sqrt{1+\frac{\U^2}{2\V^2}}\right)^{-1}$
which is independent of the momentum $p$. Although we cannot
calculate the term of order $\frac{1}{\beta^2 \V}$ in Eq.
\ref{Eq:k1} exactly, we can generally expect its functional form
to be $ \frac{\omega_n^2}{\V^2}(c_1 + c_2 p^2)$ in the small
momentum limit ($p\ll 1$). Here, $c_1$ and $c_2$ are both
constants depending on $\U/\V$. In particular, $c_2$ is
proportional to $\U^2/\V^2$ because the momentum dependence of
$k_{n,p}^{(1)}$ originates from the first order perturbation that
is proportional to $\alpha -1$ (that is proportional to
$\U^2/\V^2$). In contrast, $c_1$ should stays finite even when
$\U=0$\citep{MaldacenaStanford2016}, that is when different SYK
clusters are completely decoupled. Having established these
expectations of the terms of order$\frac{1}{\beta^2 \V^2}$ in
$k_{n,p}^{(1)}$, and keeping only the lowest order $p$ and
$\omega_n$ dependent terms, we can write
%The leading momentum dependent term must come with the second order $O\Big(\frac{1}{\beta^2 \V^2}\Big)$:
\beqn (1- k_{n,p})^{-1} = \left( \frac{3\omega_n}{2\pi\V}
\sqrt{1+\frac{\U^2}{2\V^2}} + c_2 \frac{\omega_n^2}{\V^2} p^2 +
\cdots \right)^{-1}. \eeqn Therefore, the diffusion constant,
which is proportional to the coefficient of the $p^2$ term,
vanishes in the infrared limit.

\section{Effective Action and Scaling}

The calculation above is actually consistent with a qualitative
scaling argument. As was shown in
Ref.~\onlinecite{MaldacenaStanford2016,Gu2016}, the main
fluctuation above the solution of the saddle point equations is
the ``quasi Goldstone" mode $\epsilon(\tau)$ defined as an
infinitesimal time-reparametrization $\tau \rightarrow \tau +
\epsilon(\tau)$. This reparametrization mode is also responsible
for the energy transport of the coupled SYK$_4$
clusters~\cite{Gu2016}. Its dynamics is captured by an effective
action~\footnote{As was discussed in
Ref.~\onlinecite{MaldacenaStanford2016}, in Eq.~\ref{action} the
modes with $n = 0, \pm 1$ are SL(2,R) gauged modes, and hence
should be excluded from the sum.}: \beqn \mathcal{S}_{\epsilon}
\sim \sum_{n \neq 0, \pm 1} \sum_p A \omega^2_n \left( \omega^2_n
- \left( \frac{2\pi}{\beta} \right)^2 \right) |\epsilon_{n,p}|^2 +
B p^2 |\omega_n|^m \left( \omega_n^2 - \left( \frac{2\pi}{\beta}
\right)^2 \right) |\epsilon_{n,p}|^2. \label{action}\eeqn

If the entire coupled system had an exact global reparametrization
symmetry, then the effective Lagrangian should be zero when
momentum $p=0$, because making a global uniform reparametrization
transformation on the system should not change the effective
Lagrangian, $i.e.$ the reparametrization Goldstone mode is a real
Goldstone mode. In this case the only contribution to the action
can come from finite momentum, $i.e.$ when the system makes
spatial dependent reparametrization transformation. The first
momentum independent term of the action Eq.~\ref{action} mainly
comes from the $ - i \omega$ term in $G(i\omega)^{-1}$, which
explicitly breaks the reparametrization symmetry, and hence makes
a nonzero contribution to the action of $\epsilon(\tau)$.

The second term in Eq.~\ref{action} depends on momentum $p$, which
can only come from the inter-site coupling. Because the inter-site
coupling $U^{s,s+1}$ is completely random, the leading
contribution to the $p^2$ term must be proportional to $U^2$.
Based on its definition, $\epsilon(\tau)$ has the scaling
dimension of time. In the models studied in
Ref.~\onlinecite{Gu2016,Davison2016}, the inter-site coupling is
marginal if evaluated at the on-site SYK$_4$ non-fermi liquid
ground state, thus $m = 1$ in Eq.~\ref{action}, and $B$ has
scaling dimension $[B] = 0$. The diffusion constant $D \sim B/A$
is finite. While in our case, since the inter-site coupling
$U^{s,s+1}$ has scaling dimension $-1$ at the conformal saddle
point solution, $B \sim U^2$ must have scaling dimension $-2$.
This implies that in our case $m = 3$ in action Eq.~\ref{action},
which is consistent with the calculation of $(1- k_{n,p})^{-1}$
the last section.

Had we switched the role of the two terms in $H_{q}$
(Eq.~\ref{Eq:Ham_Clust_q_2q}), $i.e.$ had we considered a toy
model with an onsite $2q$-body interaction and inter-cluster
$q-$body interaction, then the fermion Green's function would be
completely dominated by the inter-cluster interaction, and the
onsite interaction becomes irrelevant. In this case the physics
should return to the case studied in
Ref.~\onlinecite{Gu2016,Davison2016}, hence the model becomes a
diffusive metal in the infrared, and also $m = 1$ in
Eq.~\ref{action}. A similar large$-q$ calculation as we have done
in the last section confirms this expectation (see the appendix).

%\section{An ``opposite" model}

\section{Discussion}

In this work we presented a model that goes through a continuous
MIT by tuning one interaction term in the Hamiltonian. The
transition actually already occurs on a single cluster of our
model~\cite{SYKinstable}, as the infinite degrees of freedom on
each cluster support a real quantum phase transition on this zero
dimensional system. The metallic phase of this model is
essentially SYK$_4$ clusters connected by a marginal inter-cluster
four fermion coupling which obeys an independent Gaussian
distribution~\cite{Gu2016}; the insulator phase is effectively
equivalent to SYK$_2$ clusters coupled by irrelevant random
four-fermion interactions, while the on-site SYK$_2$ term is
generated spontaneously through another four fermion interaction.

In addition to time-reversal, our model Eq.~\ref{chain} also has a
local $Z_2$ fermion parity symmetry: $\chi^s_j \rightarrow \eta^s
\chi^s_j$, where $\eta^s = \pm 1$. This symmetry forbids a
inter-cluster two-body interaction. If this local fermion parity
symmetry is broken (either explicitly or spontaneously), an
inter-site (random) two-body interaction can potentially be
generated. With finite number of degrees of freedom on each site,
a two-body Hamiltonian with random hopping should lead to Anderson
localization in one dimension. However, if we take the large$-N$
limit before taking the thermodynamics limit, the conductance may
still remain finite, but this would be a large$-N$ artifact.

While completing this work, the authors became aware of another
work on MIT based on connected SYK model clusters~\cite{yaomit},
where like the model in Ref.~\onlinecite{Banerjee2016}, the tuning
parameter of the phase diagram is the ratio between the degrees of
freedoms of two inequivalent clusters of interacting fermions. It
is a very different from the mechanism of phase transition
presented in our current work.

Zhen Bi and Cenke Xu are supported by the David and Lucile Packard
Foundation and NSF Grant No. DMR-1151208. Chao-Ming Jian's
research at KITP is supported a fellowship from the Gordon and
Betty Moore Foundation (Grant 4304). The authors thank David Huse
and Juan Maldacena for very helpful discussions.

\appendix

\section{An ``opposite" model}

In this appendix, we discuss an ``opposite" model where the
inter-cluster coupling dominates the low energy physics and the
on-site SYK interaction becomes irrelevant in the IR. The model is
given by a 1D chain of SYK clusters with SYK$_{4q}$ intra-cluster
coupling and ``SYK$_{2q}$" inter-cluster couplings: \beqn H_q'=
(i)^{2q} \sum_s \sum_{1\leq i_1 < i_2 ... < i_{4q} \leq N}
V^s_{i_1 i_2 ...i_q} \chi^s_{i_1} \chi^s_{i_2} ... \chi^s_{i_{4q}}
+ (i)^{q} \sum_s \sum_{   1\leq i_1 < i_2 ... < i_{q} \leq N \atop
1\leq i_{q+1} < i_{q+2} ... < i_{2q} \leq N} U^{s,s+1}_{i_1 i_2
...i_{2q}} \chi^s_{i_1} \chi^s_{i_2} ... \chi^s_{i_q}
\chi^{s+1}_{i_{q+1}} \chi^{s+1}_{i_{q+2}} ... \chi^{s+1}_{i_{2q}}.
\label{Eq:Ham_Clust_4q_2q} \eeqn with the couplings drawn from
Gaussian distributions with zero mean: \beqn
\overline{\left(V^s_{i_1 i_2 ...i_{4q}}\right)^2} &= \frac{V^2
(4q-1)! }{N^{4q-1}}
\\
\overline{\left(U^{s,s+1}_{i_1 i_2 ...i_{2q}}\right)^2} &=
\frac{U^2 (q-1)!q!}{N^{2q-1}} \eeqn Similar to the model in Eq.
\ref{Eq:Ham_Clust_q_2q}, we will also solve $H_q'$ in the
large-$q$ limit\cite{MaldacenaStanford2016} where the coupling
constants \beqn \V^2 \equiv \frac{4q V^2}{2^{4q-1}} ~~~~~
\text{and} ~~~~~\U^2 \equiv \frac{2q U^2}{2^{2q-1}} \eeqn are kept
fixed. In this model, we expect the SYK$_{4q}$ on-site coupling
becomes irrelevant in the IR. The low temperature physics should
be governed by the SYK$_{2q}$ intra-cluster coupling. To justify
this expectation, we first solve the saddle point equations for
the 2-point Green's function: \beqn & G_s(i\omega)^{-1} = -i\omega
- \Sigma_s(i\omega)
\\
& \Sigma_s(\tau)  = G_s(\tau)^{q-1} \Big[ V^2 G_s(\tau)^{3q}  +
U^2 G_{s-1}(\tau)^{q} + U^2 G_{s+1}(\tau)^{q} \Big]. \eeqn Using
the ansatz $G_s (\tau)  = \frac{1}{2} \sgn(\tau) \Big[ 1 +
\frac{1}{q} g_s (\tau) + ...\Big]$, we obtain the form of the self
energy in the large-$q$ limit \beqn \Sigma_s(\tau) =
\frac{\sgn(\tau)}{q} e^{g_s(\tau)} \Big[ \frac{\mathcal{V}^2}{4}
e^{3g_s(\tau)} + \frac{\mathcal{U}^2}{2} e^{g_{s-1}(\tau)} +
\frac{\mathcal{U}^2}{2}  e^{g_{s+1}(\tau)}\Big](1+...), \eeqn
where $...$ represents the higher order terms in $1/q$ which will
be neglected from now on. The solutions of the saddle point
equations satisfying the boundary conditions $g^s(0)=g^s(\beta)=0$
take the form of \beqn e^{2g_s(\tau)} = \frac{1}{\frac{4
\beta^2\mathcal{U}^2}{\pi^2 v^2} + \sqrt{\frac{16 \beta^4
\mathcal{U}^4}{\pi^4 v^4} + \frac{ \beta^2 \V^2}{\pi^2 v^2}} \cos
\left( \pi v \Big(\frac{|\tau|}{\beta}-\frac{1}{2}\Big)\right) },
\label{Eq:Self_Energy_Sec2} \eeqn with the condition that \beqn
\cos \left(\frac{\pi v}{2}\right) = - \frac{1- \frac{v^2 \pi^2}{4
\beta^2 \mathcal{U}^2}}{\sqrt{1+ \frac{\V^2}{\mathcal{U}^2}
\frac{v^2 \pi^2}{16 \beta^2 \mathcal{U}^2}}}. \eeqn where $v \in
(1, 2)$. By taking the low temperature limit $\beta \rightarrow
\infty$, we can expand $v$ in terms of $(\beta \mathcal{U})^{-1}$:
\beqn
 v \sim 2- \sqrt{ 1+ \frac{\V^2}{ 8 \mathcal{U}^2}  } \frac{2}{\beta \mathcal{U}}.
\eeqn Also, as $\beta \rightarrow \infty$, $e^{2g_s(\tau)}$
reaches it low-temperature limit: \beqn e^{2g_s(\tau)} =
\frac{1}{1+ 2 \sqrt{1+ \frac{\V^2 }{8 \mathcal{U}^2}}
\sqrt{2}\mathcal{U}|\tau| + 2\mathcal{U}^2 \tau^2 } . \eeqn Notice
that, in the long time limit, we always have $e^{2g_s(\tau)} \sim
\frac{1}{2\U^2 |\tau|^2}$, which is independent from $\V$.
Physically, it means that the IR physics is dominated by the
SYK$_{2q}$ inter-cluster couplings.

To understand the transport properties of this 1D chain given by
$H'_q$, we also need to calculate the kernel $K^{ss'} (\tau_1,
\tau_2, \tau_3, \tau_4)  $ of the 4-point
function\cite{MaldacenaStanford2016,Gu2016,Fu2016}. The kernel for
our model is given by \beqn && K^{ss'} (\tau_1, \tau_2, \tau_3,
\tau_4)
\nonumber \\
&&  = - G(\tau_{13}) G(\tau_{24})  \Big[ \Big( (4q-1)  V^2
G(\tau_{34})^{4q-2} + 2(q-1) U^2 G(\tau_{34})^{2q-2} \Big)
\delta_{ss'} + q U^2 G(\tau_{34})^{2q-2}  (\delta_{s+1,s'} +
\delta_{s-1,s'})  \Big]. \eeqn Switching to the variables $\V$ and
$\U$, also switching to the momentum space, we can write the
kernel as \beqn K_{p} (\tau_1, \tau_2, \tau_3, \tau_4) =  -
\frac{\sgn(\tau_{13})}{2} \frac{\sgn(\tau_{24})}{2} \Big[   2\V^2
e^{4g(\tau_{34})} + 2\U^2 e^{2g(\tau_{34})} (1+\cos(p)) \Big].
\eeqn Here, we've only kept the leading terms in $1/q$. Let
$k_{p}$ and $\Psi_p (\tau_1 ,\tau_2)$ denote the eigenvalue and
eigenfunction of the kernel $K_p$. By the similar analysis in Sec.
\ref{Sec:large_q}, eigenvalue equation $k_p \Psi_p = K_p \Psi_p$
can be deformed into the following form: \beqn k_{p}
\partial_{\tau_1} \partial_{\tau_2}  \Psi_p (\tau_1
,\tau_2)=-\Big[   2\V^2 e^{4g(\tau_{12})}+ 2\U^2 e^{2g(\tau_{12})}
(1+\cos(p)) \Big] \Psi_p(\tau_1,\tau_2) . \eeqn Since this
equation is invariant under simultaneous translation of the two
time variables $\tau_1$ and $\tau_2$, we can directly work with
the ansatz $\Psi_{p,n}(\tau_1, \tau_2) = e^{-i \frac{2\pi
n}{\beta} \frac{\tau_1 + \tau_2}{2} }\psi_{p,n}(\tau_{12})$ with
integer $n$. Also, to simplify the equation, we work with the
variables: \beqn & \rho = \cos \left( \pi v
\Big(\frac{|\tau|}{\beta}-\frac{1}{2}\Big)\right) = \cos\left(
\frac{v}{2}(|x| - \pi) \right),
\\
& \alpha = \sqrt{1+ \frac{\V^2}{\mathcal{U}^2} \frac{v^2 \pi^2}{16
\beta^2 \mathcal{U}^2}}. \eeqn The differential equation above can
then be written as \beqn \Bigg[ v^2\left((1-\rho^2) \frac{d^2}{d
\rho^2} -\rho \frac{d}{d \rho} \right) +\frac{n^2}{4} -
\frac{v^2}{8 k_{p,n}} \left( \frac{(1+\cos p)}{1+\alpha \rho}
+\frac{4 (\alpha^2 -1)}{(1+\alpha \rho)^2}  \right) \Bigg]
\psi_{p,n} = 0. \eeqn Again, this equation is not exactly solvable
for generic values of $\alpha$. We can only proceed via
perturbative approaches. For now, we can temporarily treat
$\alpha$ and $v$ as independent variables. We will expand this
equation around $\alpha = 1$ and treat $\alpha-1$ as a
perturbation. When $\alpha=1$, this equation is exactly
solvable\cite{MaldacenaStanford2016}. The low-lying quantized
value of $k^{(0)}_{n,p}$ is given by \beqn k_{n,p}^{(0)} & =
\frac{1+\cos p }{2}\left( 1 -\frac{3|n|}{2}\Big(1-\frac{v}{2}\Big)
+ O\Big(\Big(1-\frac{v}{2}\Big)^2\Big) \right)
\nonumber \\
& = \frac{1+\cos p }{2}\left( 1 - \frac{3|n|}{2 \beta \U} \sqrt{
1+ \frac{\V^2}{ 8 \mathcal{U}^2}  } + O\Big(\frac{1}{\beta^2
\U^2}\Big) \right). \eeqn where the superscript ``$^{(0)}$"
represents the zeroth order in $\alpha-1$. However, we notice that
$\alpha -1 \sim O\Big(\frac{1}{\beta^2 \U^2}\Big) $. So even when
we include first order correction of $\alpha-1$ to $k_{n,p}$, we
should still have \beqn k_{n,p}^{(1)} = \frac{1+\cos p }{2}\left(
1 - \frac{3|n|}{2 \beta \U} \sqrt{ 1+ \frac{\V^2}{ 8
\mathcal{U}^2}  } + O\Big(\frac{1}{\beta^2 \U^2}\Big) \right).
\eeqn We know that $(1- k_{n,p})^{-1}$ should be viewed as the
propagator of low-lying modes in the system with the integer $n$
identified with the Matsubara frequency $\omega_n$. When we expand
this expression at small momentum $p$, we can write \beqn ( 1-
k_{n,p})^{-1} \simeq \left( \frac{p^2}{4} + \frac{3 \omega_n}{4
\pi  \U} \sqrt{ 1+ \frac{\mathcal{V}^2}{ 8 \mathcal{U}^2}  }
\right)^{-1}, \eeqn which indicates a pole related to diffusion
mode with a finite diffusion constant: \beqn D = \frac{\pi \U
}{3\sqrt{1+ \frac{\mathcal{V}^2}{ 8 \mathcal{U}^2} }}, \eeqn which
indicates a diffusive metallic phase at zero temperature. This
result agrees with the expectation that the low-energy physics is
governed by the inter-cluster couplings. Interestingly, even
though the on-site SYK$_{4q}$ coupling is irrelevant in the IR,
the diffusion constant $D$ still depends on its coupling strength
$\V$ in a non-trivial way.

We note that a detailed study for the case of $q = 1$ of
Eq.~\ref{Eq:Ham_Clust_4q_2q} with complex fermion will be given in
an upcoming work Ref.~\onlinecite{balentssyk}.

\bibliography{SYK}

\end{document}